\documentclass{article}

\usepackage[english]{babel}

\usepackage[letterpaper,top=2cm,bottom=2cm,left=3cm,right=3cm,marginparwidth=1.75cm]{geometry}

\usepackage{amsmath}
\usepackage{graphicx}
\usepackage[colorlinks=true, allcolors=blue]{hyperref}
\usepackage{authblk}
\title{Science of a coffee cup: a physicist walks into a bar...}
\author[1]{Aleksi Bossart}
\author[2]{Romain Fleury}
\author[2]{Benjamin Apffel*}
\affil[1]{Swiss Science Center Technorama, 8404 Winterthur, Switzerland}
\affil[2]{Laboratory of Wave Engineering, Department of Electrical Engineering, EPFL, Lausanne
CH-1015, Switzerland.}

\date{*Corresponding author: benjamin.apffel@epfl.ch}
\begin{document}
\maketitle
\begin{abstract}
... and annoys everyone with unsolicited experiments. The present paper proposes a short pedagogical review of the various phenomena that can be observed in a coffee cup with little to no equipment. The physical domains spanned include acoustics, optics and, of course, fluid mechanics. The variety of experimental and theoretical techniques introduced throughout the paper makes it suitable for a broad audience. For each topic, we first propose an experimental realization before introducing a minimal model to explain the observations. We end each section by discussing more advanced works existing in the literature as well as related applications. We provide detailed experimental procedures and videos of the experiments that can be freely used for teaching purposes. The phenomena presented here also show remarkable efficiency as icebreakers for morning coffee in laboratories or conferences.
\end{abstract}

\tableofcontents
\newpage
\section{Foreword and scope of the paper}
Visual experiments that can be witnessed in everyday life are a good opportunity for physics teachers to catch their audience's interest. They can be used as easy-to-set-up, low-cost, and elegant demonstrations of general principles encountered in textbooks. A common way to combine theoretical lectures and experiments consists in bringing together different experiments that all illustrate the same physical feature. Here, we adopt the opposite attitude and investigate many different physical phenomena with a single experimental setup. The common denominator between all areas of physics being coffee, the choice of experimental setup naturally went toward a coffee cup filled (or not) with ordinary tea-time beverages. 

This focus on a single, rather mundane physical object and the various physical phenomena associated with it presents several advantages. First, it allows students to observe and experiment freely; there is no need to railroad them toward a particular physical theory. Instead, the teacher can build upon the students' spontaneous observations and ask them to form hypotheses, which can then be further tested and refined. Many pedagogical strategies and thought-provoking questions can be employed along the way; in this article, we focus on the underlying science as a support for teachers. Second, the choice of this very low-cost setup also allows students to experiment by themselves, removes safety issues and makes it accessible to every classroom. Last, it also offers the opportunity to connect different domains of physics, and therefore invites students to build bridges between different approaches taught in different contexts.

Of course, the physics of coffee cup has been extensively discussed, in particular for teaching purposes (see for instance \cite{higbie_notes_1974, wettlaufer_universe_2011} and many references discussed hereafter published in teaching journals). However, those works either focus on a single phenomenon, or only discuss quickly and qualitatively several of them. We propose here a more comprehensive approach, which includes quantitative analysis of experimental data and explanation from scratch of a minimal model. All the experiments are conducted using standard laboratory equipment and can easily be reproduced during a teaching session.

This work is made to be as self-contained as possible, and each section can be read separately. However, the order we chose to present the experiments is not arbitrary and aims to build up a consistent story. We first start with an empty cup (section 2-3), before filling it with various liquids (section 4 to 9). Each section is devoted to a interesting physical feature and is organized as follows: an observation of a given phenomenon is first described (videos are also available) as part of a physicist's journey in coffee cup physics. We then propose, when possible, a quantitative analysis of the experiments and a minimal model explaining the observation. We finish each section with a short discussion of related effects in other contexts or more advanced works existing in the literature. 

\newpage

\section{Optical caustics}

\textit{A physicist walks into a bar on his way to the lab.}

\begin{figure}[ht]
    \centering
    \includegraphics[width=\columnwidth]{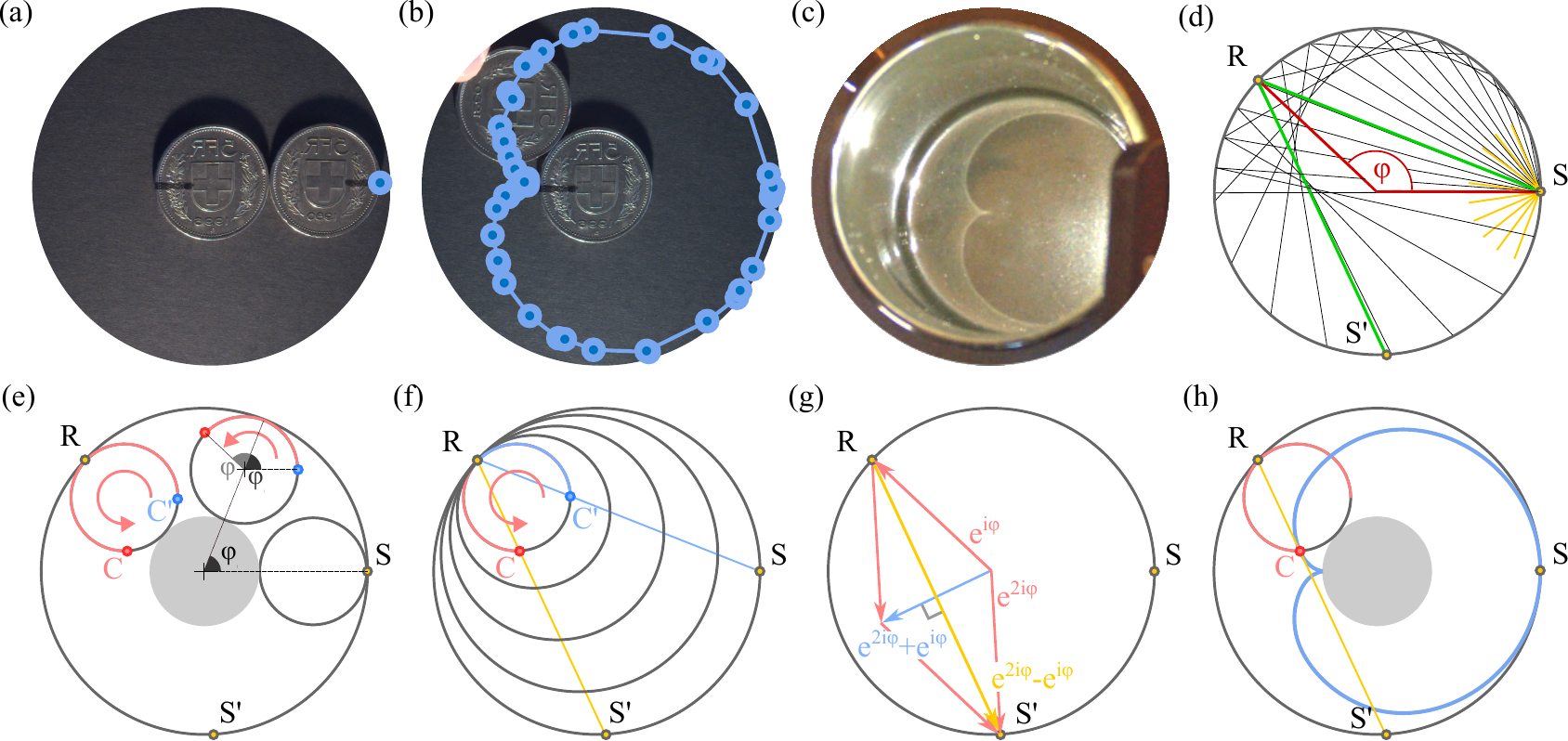}
    \caption[Caustic geometry.]{Caustic geometry. (a) Two coins. The central coin is taped in place, and a point of interest on the edge of the outer coin is indicated in blue (multimedia available online). (b) The trajectory of the point of interest as the outer coin is made to roll around the central one is drawn in blue. (c) Bottom of a cup, illuminated from the right side by a phone (multimedia available online). (d) Geometry of the situation, with light rays (green) originating from a point source $S$ located at the edge (black line) of a circular cup of radius $1$. A particular light ray is reflected at the point $R=e^{i\phi}$. The reflected ray crosses the cup edge again at $S'=e^{2i\phi}$. When all the ray are traced, the cardoid appears as a concentration of ray. (e) A disk of radius $1/3$ (small black circle) rolling around a fixed disk of the same radius (solid grey) without slipping. The $C$ point, initially located at $S$, is depicted in pink. The $C'$ point (blue) corresponds to the position that would be occupied by the $C$ point if the disk carrying it could slip but not rotate. (f) The primary light ray (blue) passes through $C'$, which implies that the reflected ray passes through $C$. (g) The tangent vector to the curve drawn by $C$ as the disk rolls is parallel to the reflected ray. (h) Cardioïd curve (blue) with the rolling disks used to define it and the reflected ray (yellow).}
    \label{fig:caustics}
\end{figure}

\textit{He takes a seat, pulls out some coins, and checks if he has enough for a coffee (Fig.\ref{fig:caustics}a). As he fidgets with a few of them, he notices something strange: when he spins one coin around another of equal size, the first coin completes two full rotations on its axis. His intuition told him it would only rotate once—after all, it only needs to travel once around the other coin’s circumference!}

\textit{Curious, he measures the trajectory of a point on the edge of the spinning coin (Fig.\ref{fig:caustics}b). It only touches the central coin once, yet it completes two full rotations. Confident he’s resolved the paradox, the physicist leans back with a satisfied sigh.}

\textit{Then he remembers: he still hasn’t ordered his coffee.}

\textit{Looking around, he sees no waiter but notices an empty cup sitting on the counter. As he peers inside, he notices a pattern of light (Fig.\ref{fig:caustics}c) resembling the path traced by the coin. Intrigued, he starts jotting down calculations on a napkin to see whether this similarity is a coincidence or not:}

\paragraph{Minimal model} To understand this phenomenon, consider the idealised situation depicted in Fig.\ref{fig:caustics}d, where a light source is placed at the edge of the cup. Focus on a single light ray: as shown in Fig.\ref{fig:caustics}d, it is reflected at $R=e^{i\phi}$. It then again meets the cup's edge at $S'=e^{2i\phi}$. The reflected ray $RS'$ can therefore be parametrised with the curve

\begin{equation}
    \rho(t, \phi)=e^{i\phi} + t(e^{2i\phi}-e^{i\phi})
\end{equation}

where $t\in[0,1]$. Now consider a disk of radius $1/3$ moving around another disk of the same radius, fixed at the center of the cup, as shown in Fig.\ref{fig:caustics}e. If the external disk can slip but not rotate, the point initially located at $S$ remains the rightmost point of the disk as it moves (point $C'$ in Fig.\ref{fig:caustics}e). On the contrary, if it could roll but not slip, it would be located on point $C$ of the same figure.

Let us now consider the light ray that goes through $C'$, as in Fig.\ref{fig:caustics}f. By construction, the reflection of $C'$ across the axis passing through $R$ and the origin lies on the reflected ray. Interestingly, since the rolling circle is a homothety of the cup's edge (Fig.\ref{fig:caustics}f), the reflected point is homothetic to $S'$. In particular, it lies precisely at an angle $2\phi$ from $C'$ on the rolling circle, exactly like the point $C$, whose trajectory we followed in Fig.\ref{fig:caustics}b. Another way to see this is to note that, by our rolling construction, we have 

\begin{equation}
    C(\phi)=\frac{2}{3}e^{i\phi}+\frac{1}{3}e^{2i\phi}=\rho(1/3, \phi),
\end{equation}
demonstrating that the reflected ray crosses the rolling curve $C(\phi)$. Furthermore, the tangent vector of the rolling curve is given by 

\begin{equation}
    \dot{C}(\phi)=\frac{2}{3}ie^{i\phi}+\frac{1}{3}2ie^{2i\phi}=\frac{2}{3}e^{i\pi/2}(e^{i\phi}+e^{2i\phi})=\frac{2\sin{\phi}}{3(1-\cos{\phi})}(e^{2i\phi}-e^{i\phi}),
\end{equation}
which is parallel to the reflected ray. This proves that the rolling curve $C(\phi)$ is tangent to the reflected ray. For a more visual proof, consider Fig.\ref{fig:caustics}g: it shows that the vectors $e^{i\phi}+e^{2i\phi}$ (blue) and $e^{2i\phi}-e^{i\phi}$ (yellow) form the diagonals of a diamond. In particular, it implies that they are perpendicular to each other. Therefore, multiplying $e^{i\phi}+e^{2i\phi}$ by $e^{i\pi/2}$ must yield a vector parallel to $e^{2i\phi}-e^{i\phi}$, \textit{i.e.} to the reflected light ray. Since this is true for every ray emanating from $S$, the $C$ curve forms an envelope against which reflected rays accumulate as in Fig. \ref{fig:caustics}d: this leads to increased light intensity, explaining the patterns we observed in our cups. 

Such an envelope is called a caustic. The particular curve we obtained by rolling coins, plotted in Fig.\ref{fig:caustics}h, is called a cardioïd \cite{castillioneus_johannes_1739}. It was originally discovered in the context of gear mechanics by R\o mer (1674) \cite{yates_curves_1974}. In the same time period, at the end of his celebrated treatise on light, Huygens studied the geometry of a optical caustic \cite{huygens_traite_1690}.

 Moving the light source around does not destroy the caustic, but slightly deforms it instead. Using catastrophe theory and increasingly refined theories of light, Berry, Nye \textit{et al.} classified caustics into a few universal types and revealed their connection to a hierarchy of three-dimensional features: wave dislocations, disclinations, and polarisation singularities \cite{nye_natural_1999}. Caustics can be seen in many places: they ripple at the bottom of clear lakes and hide in the raindrops on your windows. Even in cups, they appear in more than one way. For instance, pour hot water in your cup; you might see caustics moving around, following the convective motions of the liquid.

\section{Acoustic mode degeneracy lifting}

\begin{figure}
    \centering
    \includegraphics[width=0.9\columnwidth]{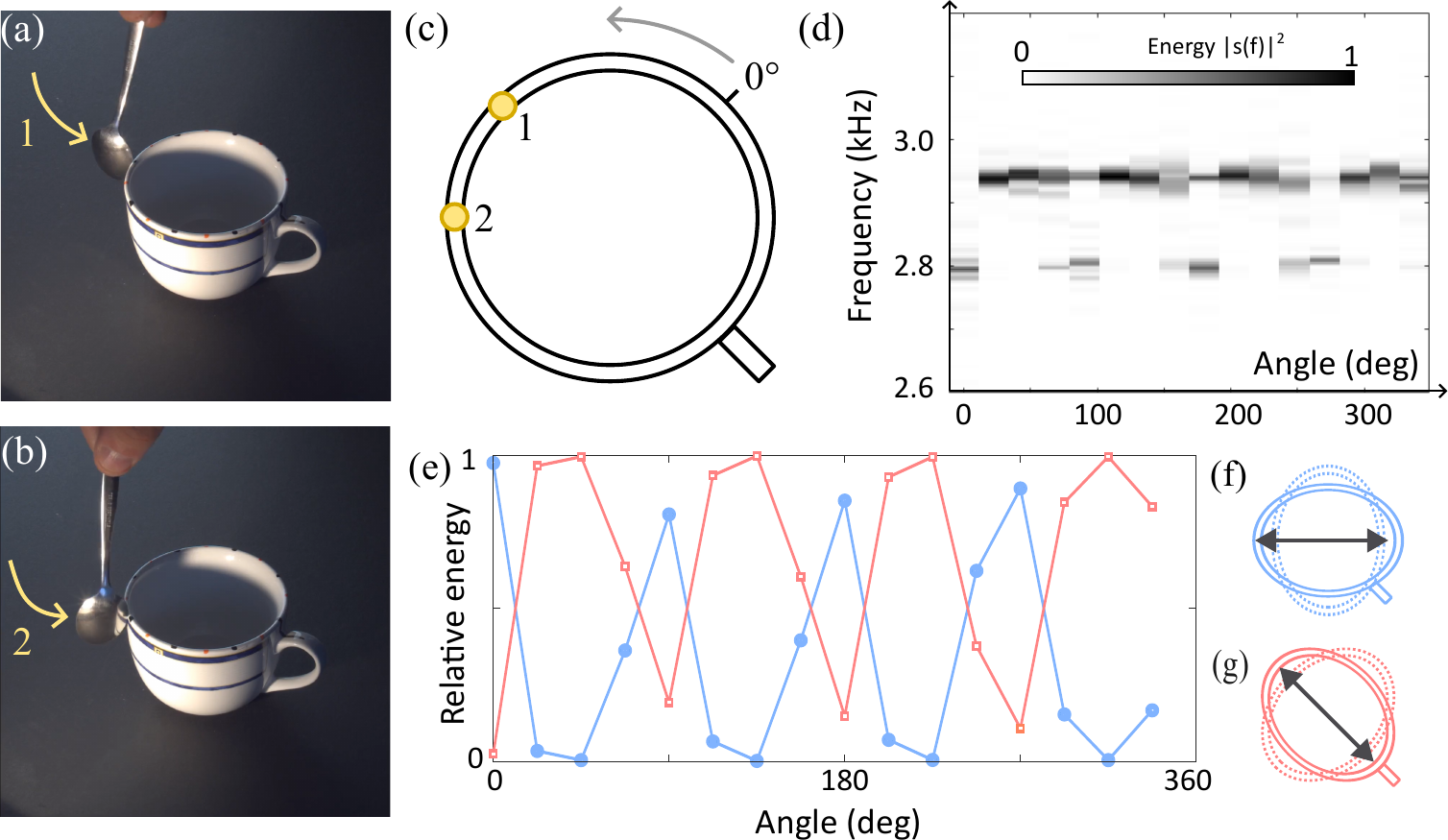}
    \caption{Acoustic mode degeneracy lifting in a coffee cup. (a-b) Hitting a cup at (a) position 1 and (b) position 2 with a spoon produces a noticeably different sound (multimedia available online). (c) Schematic top view of a coffee cup, with hit points 1 and 2 indicated in yellow. The reference $0^{\circ}$ angle is also shown. (d) Spectrogram obtained by hitting the cup in 16 locations spaced by $22.5^{\circ}$ increments, with one hitting point directly above the handle. (e) Relative energy in the two modes as a function of hitting angle. (f) First vibration mode, with the handle on a node. (g) Second vibration mode, with the handle on an antinode.}
    \label{fig:acoustics}
\end{figure}

\textit{Satisfied with his analysis, the physicist looks around for someone to share his insight. Still no one. He taps the cup with a spoon (Fig. \ref{fig:acoustics}a), hoping to attract attention. Nothing. He hits the cup again (Fig. \ref{fig:acoustics}b), and is very surprised to hear that the sound pitch was quite different from the previous impact.}
\textit{He taps the cup again in both places to confirm it: two distinct tones. Now thoroughly intrigued, he begins a systematic investigation, tapping the cup at different points and analyzing the frequencies with his phone:}

\paragraph{Minimal model}

Qualitatively, one can notice that the pitch varies with the hitting position. It is highest when we hit the cup edge at a $45^{\circ}$ angle from the cup handle and lowest at a $90^{\circ}$ angle. The notes produced also repeat when we increase angles by $90^{\circ}$ increments. This is confirmed quantitatively by the spectrograms of Fig \ref{fig:acoustics}d, which show the Fourier spectrum of the emitted sound as a function of the impact's angle. The energy is mostly concentrated in two modes with frequency $f_1 =2800$ Hz and $f_2 = 2940$ Hz, and can be computed as $E(f_i) = \int_{- \delta f}^{\delta f} |\hat s(f_i+f)|^2 df$ with $\hat s$ the Fourier transform of the sound signal and $\delta f = 40$ Hz to integrate over the whole bandwidth of each peak. From this, one can compute the relative energy densities $r_i = E(f_i)/[E(f_1)+E(f_2)]$ stored in each mode, which strongly depends on the angle $\theta$ as shown in Fig. \ref{fig:acoustics}e. 

The musical note we hear is due to the vibration of the cup's circular edge, which can be checked by pinching the edge with two fingers to absorb the sound. After an impact, a stationary wave is established, and the $90^{\circ}$ periodicity indicates that the oscillation of the cup's edge is shaped like an ellipse, as shown in Fig.\ref{fig:acoustics}(e-f). The wavelength of the deformation wave is, therefore, half of the cup's perimeter, and the vibration mode exhibits four nodes and antinodes separated by $45^{\circ}$. 

We propose to model the cup's edge by a moving mass $m$ attached to a spring of stiffness $k$, which models the restoring force due to the elasticity of the glass. The oscillation frequency is then expected to be 
\begin{equation}
    \omega_0 = \sqrt{k/m} 
    \label{eq:resAcoustic}
\end{equation}
If the cup had no handle, the location of the impact would be irrelevant due to the circular symmetry of the problem. However, the handle breaks this symmetry, which induces the frequency splitting that we observed experimentally. If we hit at $\theta=0^{\circ}$, we excite antinodes at $\theta=0 [90^{\circ}]$ while a hit at $\theta=45^{\circ}$ excites nodes at those positions (Fig. \ref{fig:acoustics}(e-f)). The handle either sits on an antinode or on a node, changing the effective mass seen by the wave in Eq. \ref{eq:resAcoustic}. From this formula, we expect the frequency to be lower when the handle is on an antinode, as the wave has more mass to carry during the vibration. This is perfectly consistent with our previous observations.  

The previous analysis can be used to estimate the relative weight between the handle and the cup from the frequency splitting that is measured. The frequency difference between the two modes is expected to be 
\begin{equation}
  f_2 - f_1 =  \frac{1}{2\pi} \left( \sqrt{\frac{k}{m}} -  \sqrt{\frac{k}{m+\delta m}} \right) \approx - f_2 \frac{\delta m}{2m}
\end{equation} where $m$ is the mass of the cup alone and $\delta m \ll m$ the added mass due to the handle. Our experimental measurement on frequencies gives $\delta m / m = 2\times 140/2940 = 9.5 \%$. A second method of measurement is needed to cross-compare the result. We could have broken our cup, but we rather chose to use Archimedes's principle to protect the cup's physical integrity. We started with a tank filled as much as possible with water (with an inverted meniscus on the edge). Holding the cup by its handle, we plunged the cup but not the handle into the tank and weighed the mass of water $m_c$  that leaked from it. We then gently dropped the cup into the tank and weighed the extra mass of water $m_h$  that went out from the tank. Assuming that the cup and the handle have the same density, one has $\delta m / m = m_h / m_c$. Two different measurements gave $\delta m/m = 9\%$ and $8\%$ respectively, in agreement with the acoustic method.

The phenomenological approach used here can be made rigorous by performing a full analysis of the cup's vibration. In such case, an equation similar to (\ref{eq:resAcoustic}) will be found with $m$ and $k$ the modal mass and stiffness of the mode, that depends as expected on the spatial distribution of the vibration \cite{richardson_modal_2000}. The degeneracy lifting, and more generally the modification of acoustic properties due to localized defects, is nowadays used routinely in civil engineering or manufacturing to perform non-invasive control of elastic structures \cite{adams_vibration_1978,fan_vibration-based_2011,hou_review_2021}.

\section{Hot chocolate/coffee effect}

\textit{The physicist happily concludes that the difference in pitch arises from the "lifting of a spectral degeneracy", dependent on the handle’s weight relative to the cup.}

\begin{figure}
    \centering
    \includegraphics[width=\columnwidth]{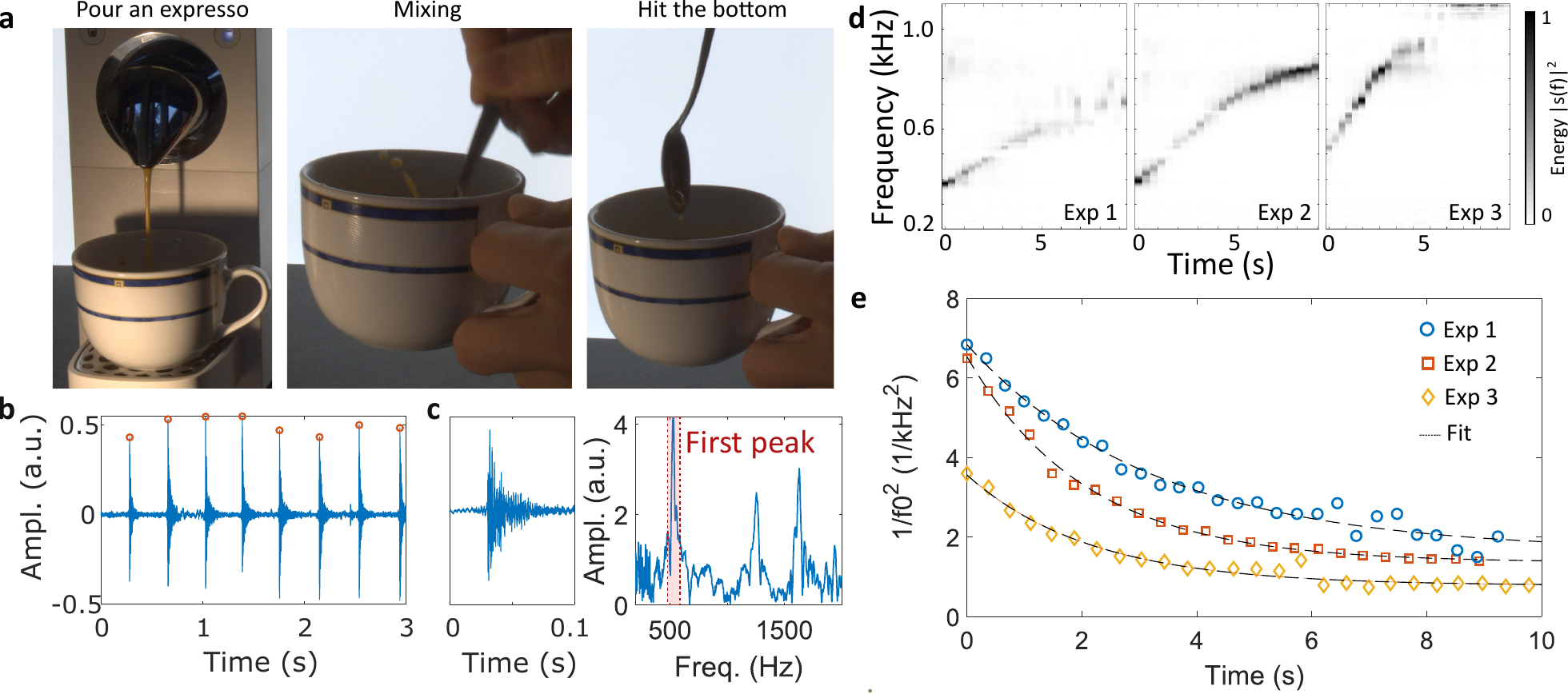}
    \caption{Hot chocolate (or here, with coffee) effect. After pouring an expresso, we mix the foam with the liquid and hit the bottom of the cup. The pitch of the generated sound rises along time (multimedia available online). (b) Sound signal recorded, each peak corresponding to a single impact of the spoon. (c) Temporal and frequency signal of the first impact. The Fourier transform exhibits several sharp peak. (d) Spectrogram obtained by stacking horizontally the spectrum of the consecutive impacts. The frequency rise of the fundamental peak along time is clear for all the experiments performed. (e) Concentration of air bubbles obtained from previous measurements and Wood's formula (\ref{eq:Wood}). A fit $f(t)=a+b^{-t/\tau}$ is performed on each curve.}
    \label{fig:hotCoffee}
\end{figure}

\textit{At this point, our physicist notices a server standing right across him, with an increasingly annoyed expression after five solid minutes of tinkling sounds. Sheepishly, he orders an espresso (Fig.\ref{fig:hotCoffee}a).
}
\textit{When it arrives, he gives it a stir. In doing so, he taps the bottom of the cup with his spoon and hears a hollow sound, quite different from before. He repeats it and hears the pitch rise slightly.
}
\textit{He tests it several times, confirming his observation, and records a spectrogram with his phone:
}

\paragraph{Minimal model} The pitch variation is induced by the injection of air bubbles in the coffee when one mixes the foam. This can be checked by noticing that the effect does not appear if one performs the same experiment with water alone.  Contrary to the previous experiment, we hit the bottom of the mug instead of the side. As a consequence, what is heard is not the vibration of the cup's edge but the stationary mode in the liquid bulk. The first mode's wavelength should be four times the height of the mug. Indeed, one imposes no longitudinal displacement at the bottom of the liquid, and constant atmospheric pressure at the top. Since pressure and displacement are in phase quadrature ($\partial_t v = - \partial_x P$), there is a pressure node at the top and an antinode at the the bottom. The first mode's frequency is then expected to be 
\begin{equation}
    f_0= c \lambda = \frac{h}{4} \sqrt{\frac{K}{\rho}}
\end{equation}
where $\rho$ is the liquid's density and $K=-V \left( \frac{\partial P}{\partial V}\right)_s$ its compressibility, which is the proportionality constant between a pressure variation $dP$ and a relative volume change $dV / V$ during an isentropic transformation. We can now consider what happens in a mixture of air and water. Following the demonstration from \cite{crawford_hot_1982}, we take subscript $a$ for the air, $w$ for water, and no subscript for total quantities. To simplify the equations, we will assume that there is a small amount of air in the water. The total volume is $V = V_a + V_w = V_w(1 + x_a)$ where $x_a = V_a/V_w \approx V_a/V$ is the volume fraction of air. One can then write $dV=dV_w+dV_a =-V_w (1/K_w + x_a/K_a) dP$ so that the compressibility of the mixture is
\begin{equation}
    K = \frac{1+x_a}{1 + \frac{K_w}{K_a}x_a} K_w
\end{equation}
We can now insert some orders of magnitude in the previous result. One has $K_w/K_a \sim 1.5 \times 10^4$, so the variation of the numerator is negligible compared to the denominator's one. With the same approximation on the density $\rho = \rho_w (1-x_a) \approx \rho_w$, one finally gets the resonance frequency known as Wood's formula,
\begin{equation}
    f_0 =\frac{c_w h}{4} \frac{1}{\sqrt{1 + 1.5 \times 10^4 x_a}}
    \label{eq:Wood}
\end{equation}
where $c_w h /4$ is simply the resonance frequency in the absence of bubbles. The previous computation shows that adding air bubbles changes the liquid's density (a bit) and the liquid compressibility (a lot). The ratio between the two variations is $K_w/K_a \gg 1$, which explains why we can essentially focus on compressibility. The later drastically changes the speed of sound, and in return the resonance frequency of the liquid column.

Inverting the previous formula, we can express the fraction of bubbles as
\begin{equation}
   \frac{1 + 1.5 \times 10^4 x_a}{f_{max}^2} = \frac{1}{f_0(t)^2}
\end{equation}
which is plotted in Fig. \ref{fig:hotCoffee}. Each experiment is reasonably compatible with a fit $1/f_0(t)^2 = a + b_0e^{-t/\tau}$, and the found values of fitting parameters give $f_{max}=[0.8;0.86;1.1] \pm 0.1$ kHz, $\tau = [3.3;2.1;2.2] \pm 0.5 s$, the uncertainty being the 95 \% confidence interval. This indicate that the fraction of air along time is exponential, which is similar to what occurs in liquid foams such as beer froth \cite{leike_demonstration_2001}. Such fitting can nevertheless be criticized in several ways. First, the maximal frequency $f_{max}^2$ varies along the experiments, while we expect it to be constant as it only depends on the amount of liquid in the cup. It can be considered constant at $f_{max} \approx 1.1$ kHz at the cost of slightly less good fitting of experimental data. This yields an estimate of $x_a\approx5\times10^{-4}$ for the initial volume fraction of bubbles in the coffee; this tiny volume fraction has an impressively large effect on the sounds produced. We can also notice that the lifetime of bubbles drastically decreases between the first and the second experiment. The reason for such a decrease is not known but it may be due to the decrease of foam on top of the coffee over time, which reduces the time during which bubbles may be trapped in the liquid.

Many experiments have shown that gas bubbles change the sound velocity, but also greatly increase the absorption of sound in liquids \cite{wood_textbook_1930,wilson_audible_2008,mallock_damping_1997,farrell_note_1969}. Such effects can be heard qualitatively in our experimental video. The model presented here is quite simple, as we used the compressibility of the air to model bubble's deformation regardless of their radius or shape. In case the bubbles start to resonate, the sound velocity will be impacted in a more complicated way \cite{wijngaarden_one-dimensional_1972}. The effect presented here is also at the heart of the Broadband Acoustic Resonance Dissolution Spectroscopy (BARDS) method used to probe the fraction of gas in a liquid during a chemical reaction \cite{fitzpatrick_blend_2012,fitzpatrick_principles_2012}. The acoustic measurement over time allows, for instance, to monitor powder dissolution in real-time, even in the case when several reactions occur simultaneously, thanks to the unique acoustic signature of each reaction \cite{fitzpatrick_blend_2012}.

\section{Tea leaf paradox}

\begin{figure}
    \centering
    \includegraphics[width=14cm]{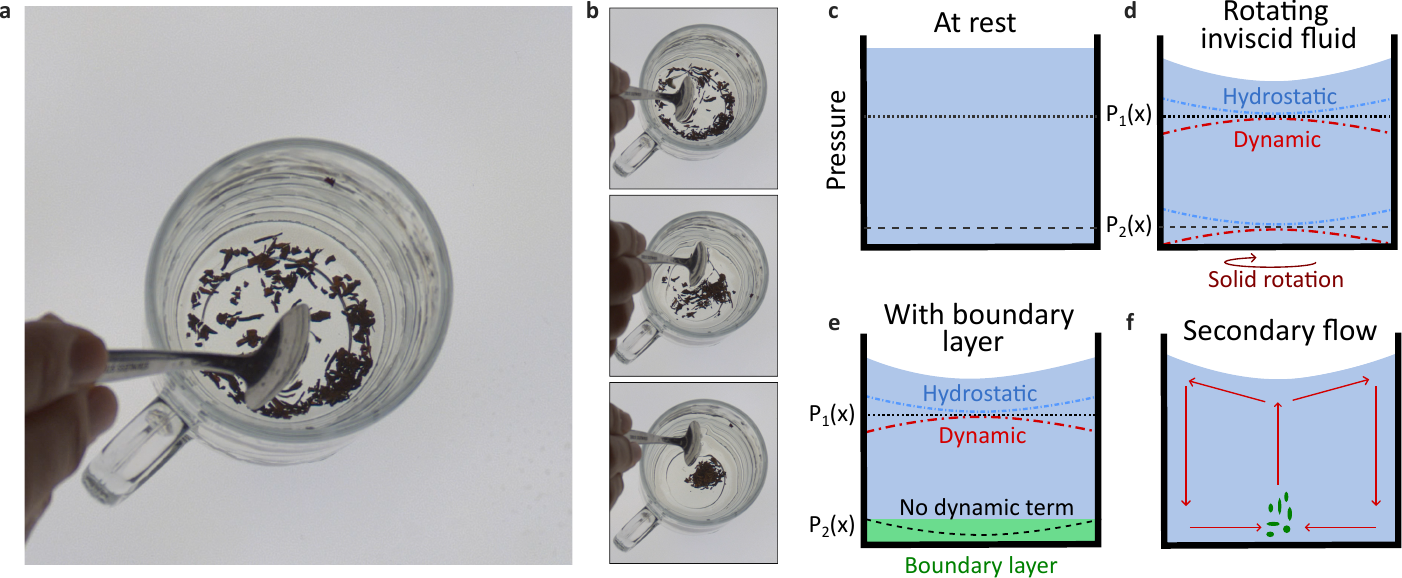}
    \caption{Tea-leaf paradox. (a) Experimental realization of tea-leaf paradox (multimedia available online). (b) Pictures from the experimental movie at $t=0$ s, $t=10$ s and $t=20$ s showing the aggregation of tea leaves at the bottom center.  (c-f) Isopressure profile $P(x) = P_i$ with $P_1 < P_2$. (c) At rest, only the hydrostatic term remains. (b) Under solid rotation of the full tank, Bernouilli's term is compensated by an extra hydrostatic term generated by the interface's bending, and the isopressure remain horizontal. (e) If the liquid is set in rotation without the tank, a boundary layer remain attached to the bottom of the tank. In this layer, the isopressure bends so that a particle placed at the bottom gets pushed toward the center. (f) Due to mass conservation in the fluid, a circulation in the vertical direction occurs.}
    \label{fig:teaLeaf}
\end{figure}

\textit{After spilling a fair amount of coffee on the counter and sounding his cup for several minutes, he catches the increasingly menacing glare of the server.
}
\textit{Growing uneasy, he decides to order a second drink to placate the server. In need of inspiration, he glances at his counter neighbor, who has ordered tea and is stirring it thoughtfully. He watches as the tea leaves dance at the bottom of the cup and eventually all gather at the center of the cup (Fig. \ref{fig:teaLeaf}a-b)}

\paragraph{Minimal model} At first sight, the experiment is really puzzling for the intuition. Since tea leaves are initially at the bottom of the tea cup, it means that their density is higher than the one of the surrounding water. Applying a centrifugal force to the liquid should therefore eject them on the outside of the cup. Solving this paradox requires to perform a full analysis of the flow that is generated by the moving spoon. 

When the liquid is at rest, both the interface and isobars $P(x) =P_i$ are horizontal, as shown in Fig. \ref{fig:teaLeaf}c. If the full tank is now set in rotation with angular velocity $\Omega$, the liquid will, after some time, eventually reach a solid rotation regime. Considering the fluid in the rotating frame and assuming that everything is invariant under rotation, the pressure field $P(r, z)$ verifies
\begin{equation}
    \vec \nabla P = \rho (\vec g - \vec a)
\end{equation}
where $\vec a = - \Omega^2 \vec r$ is the acceleration of the fluid at this location. The previous equation gives $ \partial_r P = \rho r\Omega^2$ so that $P = f(z) + \frac{1}{2}\rho r^2 \Omega^2$, and $\partial_z P = -\rho g$ so that $P = F(r) -\rho g z$. Combining those equations, one can determine the functions $f(z)$ and $F(r)$, and the pressure field is finally
\begin{equation}
    P(r, z) = P_0 + \frac{1}{2}\rho \Omega^2 r^2 - \rho g z
\end{equation}
We recover some kind of Bernoulli relation, but the sign in front of the dynamical term is inverted compared to the usual case. Indeed, if one applies naively Bernouill's principe, the pressure expresses as $P = C - \frac{1}{2}\rho \Omega^2 r^2 - \rho g z$ with $C$ a constant, which contradicts the previous equation. The resolution of this apparent paradox lies in the fact the pressure in Bernouilli's principle is computed along a given streamline. In particular, the (non-)constant $C$ depends on the circular streamline that is considered in our rotating fluid. If the flow is irrotational, the constant is the same for all the streamlines and Bernouilli's relation is thus often used to compute pressure field globally in this context. The present flow being nevertheless heavily rotational, this method cannot be applied here.

The surface of the fluid being in contact with the air, it must be an isobar area $P(r,z)=P_{atm}$. The height profile $z(r)$ is thus given by 
\begin{equation}
    z(r)=z_0 + \frac{\Omega^2 r^2}{2g} 
\end{equation}
and we recover the fact that the interface of a rotating liquid is parabolic. Alternatively, the previous equation can also be obtained by equaling the centrifugal force and the weight at the interface in the rotating frame. \cite{berg_rotating_1990}

Let us now imagine that we replace a fluid particle by a solid one of the same density. In this case, the centrifugal force (or equivalently, the dynamical term in the pressure field) is compensated by the hydrostatic pressure term due to the deformed interface, and no motion should arise (Fig. \ref{fig:teaLeaf}d). This is true in the bulk of liquid, but the tea leaves are concentrated at the bottom of the cup. Since they are heavier than the fluid particles they replace, they should still be pushed toward the outside as the centrifugal force acting on them overcomes the horizontal pressure gradient. At this stage, the paradox therefore remains unsolved. 

There is, however, a big difference between rotating the tank with the liquid or rotating the liquid only. In the latter case, the edges of the tank (including the bottom) remain still while the liquid bulk performs solid rotation. If one imposes a no-slip boundary condition, there must therefore exist a transition between the still liquid attached to the bottom and the rotating liquid bulk. This transition occurs on a small area called a boundary layer, which physically corresponds to the small portion of space where viscous effects cannot be neglected. Its vertical size $\delta$ can be estimated from dimensional analysis. Assuming it depends on the rotation speed and the kinematic viscosity only, the length scale that can be built from those quantities is $\delta \sim \sqrt{\nu/\Omega} \sim 1$ mm. Such boundary layer appears similarly in van Karman flow near a spinning disk, for which analytical computation gives similar expression for the boundary layer's size \cite{karman_uber_1921}.

As the velocity is zero at the bottom to fulfill the no-slip boundary condition, only the hydrostatic term remains and the isobar $P_2(x)$ in the boundary layer bends toward the outside of the cup (Fig. \ref{fig:teaLeaf}e). At the bottom of the cup, the pressure is thus higher on the outside of the cup than near the center. The resulting pressure gradient will push the fluid particles at the bottom toward the cup's center. As water is incompressible, those particles will be pushed by the one following them, and an ascending current will be established. Following the path, a global rolling circulation eventually establishes in the fluid as sketched in Fig. \ref{fig:teaLeaf}f. The tea leaves being heavier than the fluid, they cannot follow the ascending current and remain trapped at the bottom center where the ascent starts. On the other hand, floating particles at the surface will be pushed near the edges, as shown by one of the floating leaves in our video.

This phenomenon is an example of bulk transport induced by boundary layers effects, which occur in many different contexts. For instance, it explains why sediments aggregate on the inward side of river's meanders \cite{bowker_albert_2007,einstein_ursache_1926,thomson_v_1997,leopold_river_1960}. It also explains some atmospheric circulations between low and high-pressure areas \cite{tandon_einsteins_2010}. They are also used in laboratories to aggregate particles for manufacturing purposes \cite{zhang_einsteins_2023} or in blood-plasma separation \cite{yeo_electric_2006}. Note that the use of a 'deformable spoon' was also recently investigated in the tea-leaf paradox and the stirrer-flow coupling exhibits rich phenomenology \cite{apffel2024tealeaf}.

\section{Double diffusion convection}

\begin{figure}
    \centering
    \includegraphics[width=15cm]{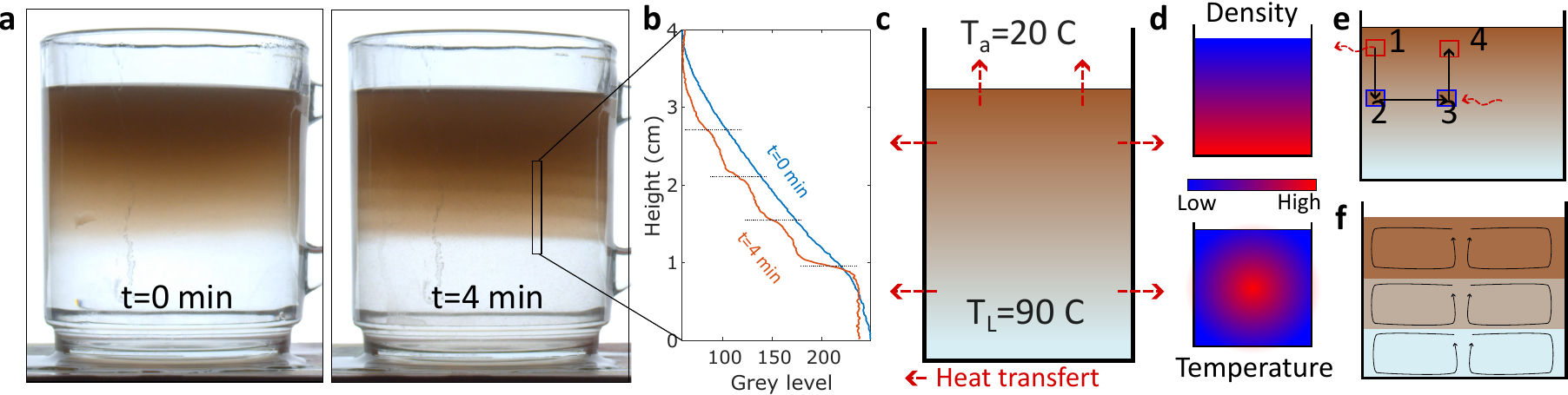}
    \caption{Layering of a latte. (a) By dropping a shot of hot chocolate into hot salted water, a density gradient encoded in the color establishes. After a couple of minutes, discrete horizontal color steps appear (multimedia available online). (b) Grey level of each image in the vertical direction and averaged over 20 pixels in the horizontal one. Starting from continuous variation at $t=0$ min, the discrete step appear for $t=4$ min. (c) Sketch of the experiment showing the initial density gradient and thermal exchanges (red arrow). (d) Sketch of the density and temperature profile in the cup. (e) Motion of a fluid particle in the cup. (1) A particle at the edge cools down near the cup's edge, which increases its density. It therefore starts to sink, and eventually reaches an area of higher density (2). As it cannot sink anymore, it gets pushed horizontally by the particles following it (3). The particle being back to the center, it heats up gain and rises as its density decreases (4). (f) Those circulations perform mixing within horizontal layers, explaining the appearance of discrete color layers.}
    \label{fig:latte}
\end{figure}

\textit{Feeling a little dizzy from looking at vortices for so long, and made uncomfortable by the inquiring look of his tea-mixer neighbor, he decides to skip the tea and opts for a latte.
}
\textit{The waiter, showing evident expertise, prepares a beautifully continuously layered latte, going from dark brown on the top to full white at the bottom (Fig. \ref{fig:latte}a). After a couple of minutes, however, the color layered turned discrete, showing sharp edges between the different color areas:
}

\paragraph{Minimal model} Due to its initial velocity, the espresso penetrates in the denser milk (or, to ease the experiment, salty water as in Ref. \cite{xue_laboratory_2017}) until buoyancy stops it. After a short transient regime, the mixture becomes darker at the top than at the bottom (Fig. \ref{fig:latte}a), which indicates that there is more dyed water, and so less salty water, at the top of the cup. A darker region is thus associated with a smaller liquid density, and the color distribution directly indicates the vertical density stratification in the fluid. However, there also exists a temperature difference between the center of the cup and the edges due to thermal exchanges (Fig. \ref{fig:latte}c). Due to the temperature gradient, the liquid at a given height is denser at the edges than in the center. There are therefore two gradients in the fluid: a vertical density gradient and a horizontal temperature gradient that also impacts the density profile. 

Due to molecular and thermal diffusion, both quantities will eventually become homogeneous if one waits long enough. However, the typical time scale at which those processes occur is not the same. The ratio between the two can be estimated by comparing the diffusivity constants, which are $D_T \sim 10^{-3}$ cm$^2$/s for temperature and $D_c \sim 10^{-5}$ cm$^2$/s for salt diffusion in water. From this, one can deduce that the diffusion of salt is $D_T/D_c \sim 100$ times slower than thermal diffusion, showing that the density gradient will remain much longer (typically several days) than the temperature one (which disappears after typically one hour).

We can now consider what happens to a fluid particle in the cup, as in Fig. \ref{fig:latte}e and in Ref. \cite{mendenhall_stratified_1923}. If the fluid particle is initially located at the edge, it will cool down due to thermal transfer with the outside. As it does so, its density becomes higher than that of the surrounding fluid, meaning that it starts to sink. If the mixture were uniform (same color everywhere), it would sink until the bottom of the cup and create a large convective cell in the cup. However, there exists a density gradient in the fluid, so that the particle cannot go beyond a certain depth due to buoyancy. But since other fluid particles are following it, our fluid particle must now perform horizontal motion along this effective bottom to leave space for the incoming ones. When it comes closer to the center, the particle heats up again, its density decreases and it start to rise again. The horizontal temperature therefore generates large circular flows that are called convection cells (Fig. \ref{fig:latte}f). Due to the vertical density gradient, several horizontal convection cells appear on top of each other in the cup. The generated flow performs mixing within the cell, which explains the appearance of uniformly colored layers with discontinuity between each. The color discontinuity materializes the separation between two adjacent convection cells. 

The existence of time scale separation between thermal and salt diffusion is crucial for the previous explanation to hold, as it allowed to consider a fluid particle of constant salt concentration during a full thermal cycle in Fig. \ref{fig:latte}e. If salt diffusion was faster than thermal diffusion, the particle would sink until the bottom by changing its salt concentration along its trajectory, and the discrete layer would not appear. We can estimate the size of a horizontal layer in the following way. During cool down, the density varies by $\Delta \rho_T = \rho_w \alpha\Delta T$, where $\rho_w =1.0$ kg/L is the water density, $\alpha = 2.6 \times 10^{-4} {}^\circ$C$^{-1}$ is the thermal dilatation of water, and $\Delta T \sim 50 ^\circ$C is the temperature drop during cool down. On the other hand, the spatial density profile due to varying salt concentration can be estimated from the color profile. The salty water had salt concentration of about $0.1$ kg/L, and thus a density $\rho_s \approx 1.1$ kg/L, while the colored water has the same density as water $\rho_w = 1.0$ kg/L. From Fig. \ref{fig:latte}b, we see that the color gradient is approximately linear and goes from fully white (no colored water) to fully brown (only colored water) on a length $L \approx 4$ cm. Along the color gradient, going down by a length $d$ thus leads to density variation of $\Delta \rho _s \approx (\rho_s - \rho_w) \times d/L$. As explained before, a cooled down, less salted liquid volume will stop to sink after a distance $d$ when its density matches the one of the hotter, more salted liquid surrounding it. Along the vertical direction of a horizontal layer, both density variation should therefore be equal $\Delta \rho _s = \Delta \rho_T$. From this, we recover $d = \alpha L \rho_w \Delta T / (\rho_s-\rho_w) \approx 5$ mm. This is remarkably close from our experimental observation on Fig. \ref{fig:latte}b where a layer has a typical size of $d \approx 6-7$ mm. Alternatively, this method can also be used to estimate the thermal dilatation coefficient from the observed height $d$ of mixing cells, which is found to be $\alpha \approx 3 \times 10^{-4} {}^\circ$C$^{-1}$ in reasonable agreement with the exact value.

Such phenomena, named ‘double diffusion convection’, not only occurs in latte \cite{xue_laboratory_2017,chong_cafe_2020} but also in the oceans \cite{radko_double-diffusive_nodate,schmitt_double_1994} or during iceberg melting \cite{huppert_melting_1978}, and is a decisive mixing mechanism in geophysical flows. A similar analysis can also be carried to explain the formation of salt fingers \cite{schmitt_oceans_1995},

\section{Coffee stains}

\begin{figure}
    \centering
    \includegraphics[width=14cm]{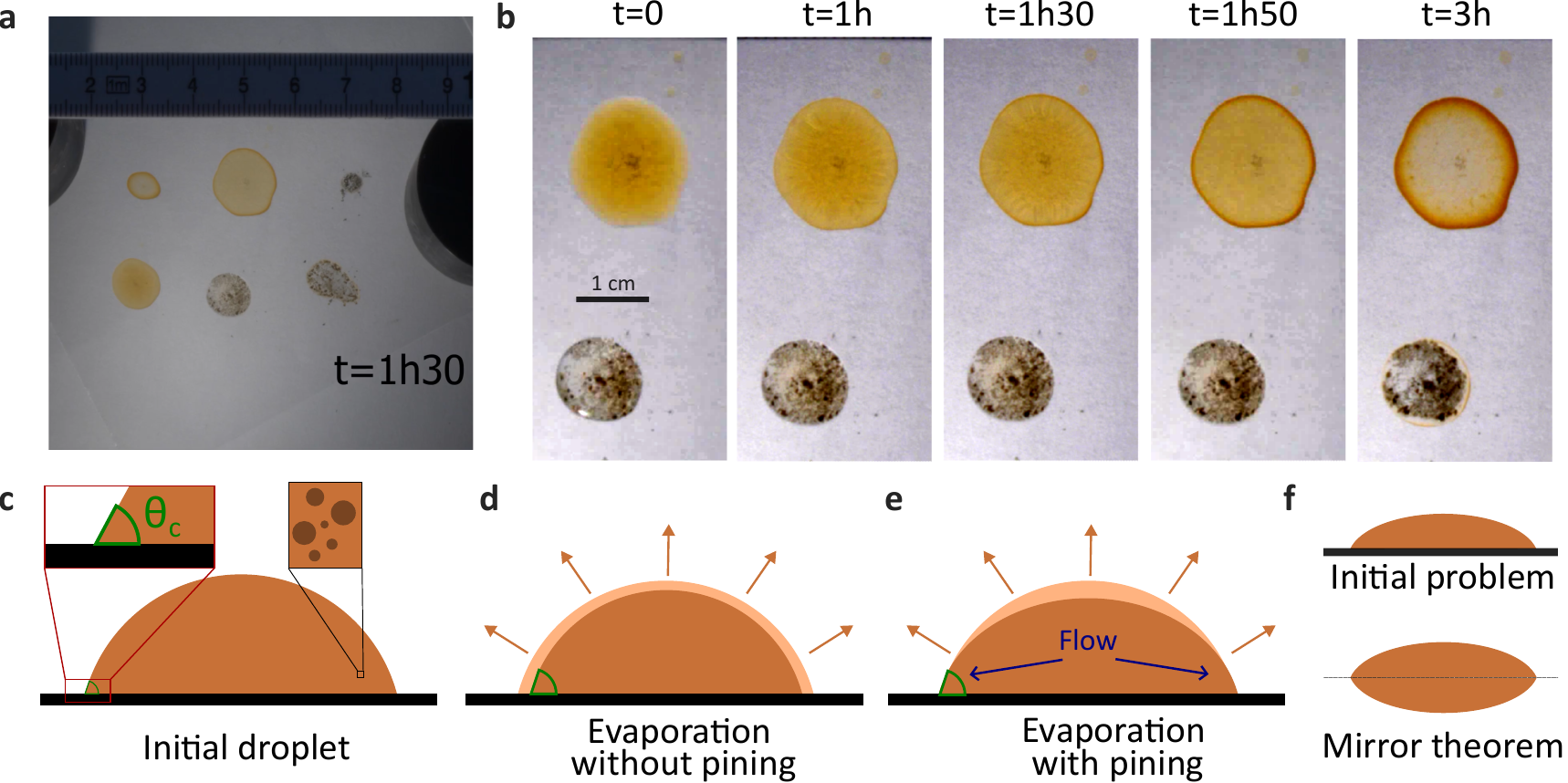}
    \caption{Formation of coffee rings. (a) Snapshot of the experimental movie with three hot coffee drops (brown colors) and three water droplets containing pepper grains (mutimedia available online). (b) Detail of the experimental movie for different times showing the difference between coffee and pepper droplets. (c) A droplet of coffee contains many coffee grains which diameter is typically 10-100 $\mu m$. The contact angle of the droplet is fixed by the choice of substrate. (d) If the contact line could move, the droplet would evaporate uniformly along time. (e) If the contact line is fixed, the liquid redistributes inside the droplet to compensate for the evaporation occurring at the edges, leading to transport of particles at the edge. (f) Equivalent problem using the mirror image theorem.}
    \label{fig:CoffeeStain}
\end{figure}

\textit{Both amazed by the formation of discrete layers and strongly disappointed by the terrible taste of this salty latte, the physicist finally decides to leave the bar, realizing that night fell and that he never made it to the lab. However, just before leaving, he notices that the coffee droplets he spilled during the mixing dried on the table. Intriguingly, these coffee stains are not uniformly brown but rather exhibit noticeably darker edges (Fig. \ref{fig:CoffeeStain}a-b):
}

\paragraph{Minimal model}
One should first describe what coffee really is from a physical point of view. Although coffee molecules are very small (about $\sim 1$ nm for caffeine), they exist in an expresso in the form of clusters of about $\sim 10-100 \mu$m diameter. Coffee, therefore, consists of a suspension of coffee grains in water. When a coffee droplet is placed on a surface, the contact angle $\theta_c$ (Fig. \ref{fig:CoffeeStain}c) at the edge of the droplet depends on the surface properties and can go from $0^\circ$ for wetting surfaces to almost $180^\circ$ for very hydrophobic ones. In our experiments, we used a PMMA plate on which coffee drops exhibit good wettability ($\theta_c < 90^\circ$). When a droplet is deposed, it starts to evaporate and the droplet should therefore shrink. Naively, one could think that evaporation occurs homogeneously and that both radius and height will diminish over time, leaving a uniform stain on the table as in Fig. \ref{fig:CoffeeStain}d.

However, this picture is wrong because the contact line of the droplet is pinned on the substrate due to the microscopic roughness of the latter. Therefore, both the radius and the contact angle of the droplet are fixed, as confirmed by the largely constant droplet size observed over the course of the evaporation in Fig.\ref{fig:CoffeeStain}b. If one still assumes uniform evaporation over the surface, then some liquid must flow toward the outside to compensate for the evaporating liquid at this location. Coffee colloids are transported with this flow toward the outside, which results in higher concentration at the edge of the droplet. Such migration was confirmed by particle imaging under a microscope \cite{deegan_capillary_1997,university_of_pennsylvania_coffee_2011}

This is still not the full picture, and a quantitative approach to the problem can be performed following Ref.  \cite{deegan_capillary_1997}. We consider the diffusion problem of water steam concentration $c(\vec x, t)$ in the air near the droplet. We will assume that a stationary regime is reached so that the diffusion equation reads $\Delta c = 0$ in the bulk. At the edges, we impose $c=c_a$ at infinity (with $c_a$ the ambient concentration), $c=c_s$ at the droplet's edges (with $c_s > c_a$ the saturating steam concentration in the air) and $\partial_{\vec n}c = 0$ at the solid interface (no flux condition). Using the mirror image theorem from electrostatics, the solid surface acts as a perfect mirror, and the problem becomes equivalent to considering a symmetrized droplet without the substrate as in Fig. \ref{fig:CoffeeStain}f. Keeping the electrostatic analogy and using linearity, we are now looking for a potential $c$ that verifies $\Delta c = 0$ in the bulk, $c= c_s-c_a$ at the droplet's edge and $c=0$ at infinity. Due to the tip's influence, one expects $\nabla c$ (\textit{i.e.} the electric field) to diverge at the angular point. This can be shown analytically using complex analysis and conformal invariance  \cite{appel_mathematics_2007}. This gradient, in our case, is the evaporation flux, which shows that the latter diverges at the contact line. The previously discussed hypothesis about uniform evaporation is quantitatively wrong, as more evaporation occurs at the edge of the drop than on its top. The flow induced by evaporation will therefore be stronger than expected, as more liquid needs to be sucked toward the outside.

The formation of circles strongly depends on the chosen experimental conditions. The surface roughness, the surface tension of the liquid, the size of the particles can for instance greatly impact the phenomena \cite{deegan_pattern_2000,yunker_coffee_2013, yunker_effects_2013} or even suppress it \cite{yunker_suppression_2011}. It can be seen qualitatively in Fig. \ref{fig:CoffeeStain}b: a water droplet filled with pepper grains much larger than coffee particles does not produces a ring-shaped pattern while evaporating. Such problem has important consequences in ink-deposition \cite{park_control_2006} or DNA chips manufacturing \cite{dugas_droplet_2005}

\section{Conclusion}
\textit{'It's crazy that I always have to serve those physicists salty water to get them out', thought the waiter as he closes the bar.}

This short review discloses the potential of a simple coffee cup to perform various physics experiments. The variety of techniques used makes it particularly suitable for experimental conferences or undergraduate lectures. We did our best to bring together the experiments that are the most spectacular or surprising from our point of view, while remaining doable in a classroom with minimal equipment. Other fascinating problems such as droplet levitation above a coffee \cite{ajaev_levitation_2021,umeki_dynamics_2015} or walk-induced coffee spilling \cite{mayer_walking_2012} were therefore not discussed here. We hope that this work will encourage experimental demonstrations in classrooms, in order to help students to connect their theoretical knowledge with concrete examples of their everyday's life. 

\section{Acknowledgment}
This paper results from  a conference talk given at 'Les Gustins Summer School' in 2024. The author therefore acknowledge all attendees for their precious feedback.
\section{Data availability and competing interest}
All the data generated during this study are available in the paper or in the supplementary movies. Additional data or codes for analysis are available from the corresponding author upon reasonable request. The authors declare no conflict of interest.

\section{Experimental troubleshooting}
We describe hereafter some troubleshooting to ease experimental reproducibility. We highly recommend to test each experiment before performing it in front of an audience.

\paragraph{Optical caustics :} for the coin experiment, it might be difficult to perform non-slipping motion. The best results were obtained by taping the central coin and push from the outside of the moving coin to increase normal, and thus tangential, stress. This experiment is also the opportunity to make some reminders about how friction stress works. Alternatively, the use of circular gears could also greatly ease the experiment. Concerning the optical experiment, best results are obtained with black flat-bottom mugs and punctual sources such as a phone's light. A pie mold also provides great visualization and may be more suitable for larger audience.

\paragraph{Acoustic mode degeneracy :} this experiment works with the vast majority of cups, but the best results are obtained with porcelain cups that provide clear tone with long sustain. The effect can be made even clearer by using a crystal wine glass and attaching a mass on it. The experimental signal was recorded with the built-in microphone of our laptop and is of good enough quality to perform quantitative analysis.

\paragraph{Hot chocolate/coffee effect :} the strongest effect was observed in our case with coffee just poured by the machine. The mixing of the foam combined with the very hot liquid ensures excellent mixing of air in water. It is very important to hit the bottom of the cup to excite the liquid's mode. We did not succeed to hear clearly the effect with hot chocolate and cold water, hot water was working better but not as well as hot poured coffee. 

\paragraph{Tea-leaf paradox:} the experiment presents no particular difficulty. A gentle mix with the spoon on the upper part of the water is enough to observe the effect. It is interesting to have both floating and sinking leaves to show the differentiated behavior depending on the location in the fluid.

\paragraph{Double diffusion convection :} this experiment requires a bit of care to be successful. The use of very hot water ($\sim 90^\circ$C) makes the phenomena easier to observe as convective effects are stronger. The key parameter is the penetration of the hot chocolate shot in the salty water. If it is too small, the dyed water remains at the surface and no clear convective cells appear. If it is too large, the mixing is homogeneous and no color (and density) gradient appears. It requires of bit of testing to obtain a nice color gradient as in the experimental video, and the two parameters that can be played with are the penetration velocity of the chocolate and the amount of salt. After typically 5 to 10 minutes, we observed in our experiment some 'condensation' of chocolate powder that started to fall at the bottom of the cup, which is very nice but slowly destroys the layers. The experiment can be also be made in principle with milk instead of salty water, but we did not manage to obtain as nice color gradients, indicating a lack of practice in latte preparation.

\paragraph{Coffee stains :} it is important to use coffee, as for instance the experiment did not work in our case using hot chocolate or pepper grains.
\bibliographystyle{abbrv}
\bibliography{biblio}

\end{document}